\documentclass{aa}

\usepackage[varg]{txfonts}
\usepackage[colorlinks=true, linkcolor=blue, citecolor=blue, filecolor=blue, urlcolor=blue]{hyperref}
\usepackage{caption}
\usepackage{subcaption}
\bibpunct{(}{)}{;}{a}{}{,} 

\begin{document}

\title{Clusters of tribocharged dust aggregates as pebbles in protoplanetary disks}

\author{F. C. Onyeagusi
\and J. Teiser
\and G. Wurm} 

\institute{University of Duisburg-Essen, Faculty of Physics,
Lotharstr. 1, 47057 Duisburg, Germany\\
\email{florence.onyeagusi@uni-due.de}}

\date{Received date /
Accepted date }


\abstract{In recent years, the tribocharging of colliding and bouncing submillimeter (submm) particles has been studied as a possible mechanism promoting the formation of large pebbles on centimeter (cm) to decimeter (dm) scales in protoplanetary disks. Here, we observe, for the first time, that it is not only monolithic, spherical particles, but also real dust aggregates, that become tribocharged and end up forming large clusters. For aggregates of $\sim 0.4$\,mm consisting of $\rm \sim$ 1\,micrometer (µm) sized dust, we determined net charge densities up to $10^{-7}$\,C/m² during our drop tower experiments. These charged aggregates form compact clusters up to 2\,cm in size via collisions with other clusters and aggregates at collision velocities on the order of 1\,cm/s. Size and speed are the only lower limits for growth, currently set by the limits of the experiment. However, these clusters already form under conditions that are well beyond the expected transition to bouncing for uncharged aggregates and clusters. Our findings further support the idea that collisional charging can leapfrog the traditional bouncing barrier and form larger clusters that then serve as large pebbles. These cm-sized clusters are more susceptible to further evolutionary steps via particle trapping, concentration, and planetesimal formation.}



\maketitle

\section{Introduction} \label{sec:intro}

Planet formation includes various mechanisms that drive the transition from one size regime to another. In between, different barriers entail detours in terms of their growth. The very early history of planetesimal formation is dominated by the collisional evolution of small micron-sized dust grains in protoplanetary disks \citep{Blum2008, WurmTeiser2021, Birnstiel2024}. Due to relative motion caused by thermal, drifting, or turbulent motion, particles collide with each other. Initially, they stick together and become fractal-like \citep{Wurm1998, Paszun2006, Schubert2024}. They restructure eventually and form compact aggregates \citep{Blum2000, Meisner2012}. Lacking ways to dissipate their kinetic energy, they then enter a phase of bouncing, also known as the bouncing barrier \citep{Zsom2010}.  

The existence of a bouncing barrier is a rather common outcome of laboratory collision experiments \citep{Kelling2014, Demirci2017, Kruss2020}. It also shows up in numerical simulations \citep{Arakawa2023Bouncing}. As this concept is not restricted to astrophysics, compact dust aggregates of a limited size range can readily be generated if dust is vibrated \citep{Weidling2009, Onyeagusi2024}. This way, simulants for dust aggregates at the bouncing barrier can be produced. The final size of these aggregates depends on the collision velocities and dust grain sizes. For micron-sized silicates, it can be tuned to the order of 1\,millimeter (mm) \citep{Onyeagusi2024}. 

In any case, collisional growth in disks gets stalled at this point. To proceed on the size scale, hydrodynamic interactions are invoked to trap and concentrate particles \citep{Raettig2015, Squire2018, Marel2023}. These mechanisms are, however, sensitive to the particle size or Stokes number and mm particles seem to be too small to trigger any instabilities \citep{Carrera2022, Drazkowska2014}. In this work, we do not enter into a discussion of the sensitivity of these instabilities or elaborate on the smallest size limit for a particle cloud that might evolve further. We only note that it might not be the best choice of models if planet formation depended on a stretch or finetuning in size between the collisional and the hydrodynamic regimes. A robust transition would let particles continue to grow well into the next evolutionary phase.

Electric charges have a great potential to support attraction between grains and allow for the growth of larger entities. There are multiple ways to generate charges on small particles and aggregates in a protoplanetary disk, such as plasmas generated by ionizing radiation or photoemission. Temperature variations or transient heating events can lead to charge separation in grains \citep{Horanyi1990}. Even without this thermally induced charge separation, secondary electron emission can generate charges of the opposite polarity on similarly sized grains of the same material, as the equilibrium charge equation can have multiple roots \citep{MeyerVernet1982}. For aggregates specifically, there is a strong dependence on the porous structure in comparison to monolithic grains when it comes to charging through plasmas or photoemission \citep{Ma2013}. In the context of planet formation, these aspects might be of special relevance in regions where external ionization sources by energetic radiation are efficient. Here, electric charging in a plasma might also be limited by repelling Coulomb forces  \citep{Okuzumi2009, Akimkin2023}. In any case, in the opaque midplane in the inner disk, plasmas are rather weak and photoemission is not significant.

An efficient way of charging with a high potential to eventually form larger particles in the midplane is tribocharging. \cite{Steinpilz2020a} showed that mm glass beads, which become charged in mutual collisions, form clusters of beads that are much larger than neutral beads could ever do. They conducted microgravity experiments and numerical simulations, agreeing that electric charge is the driving factor in particle aggregation. A number of works -- both experimental and using numerical simulations -- support this idea of a phase of cluster growth of charged grains in comparison to neutral grains bouncing \citep{Matthews2011, Lee2015, Yoshimatsu2017, Jungmann2018, Singh2018, Singh2019, Steinpilz2020b, Teiser2021, Jungmann2021c, Jungmann2022, Onyeagusi2022, Schwaak2024, Fuehrer2024}. In fact, it seems possible that one or two orders of magnitude can be bridged by forming clusters of charged grains, up to at least several cm in size.
So far, all these results have been based on experiments with solid monolithic particles, usually using monodisperse, spherical glass, or basalt particles. However, real bouncing particles in protoplanetary disks will be dust aggregates consisting of µm-sized grains. Thus, a question remains as to whether, chargewise, porous dust aggregates behave like monolithic spheres of similar size. From experiments on wind-driven erosion of planetesimals, it has recently been shown that dust aggregates behave the same way as glass beads with respect to sticking \citep{Schoenau2023}. More precisely, dust aggregates have not been found to be stickier than their glass bead analogs. 

The question of whether dust aggregates also behave similarly, that is, whether their charging is more or less effective in collisions by tribocharging is not trivial. Tribocharging is a complex process and there is no unified picture of the underlying mechanisms \citep{Lacks2019}.
For instance, it is not clear a priori whether dust aggregates are indeed strongly charged in collisions. A few other unknowns in this area include: what happens if grains shift within the aggregate and whether they charge internally to already form more stable dust aggregates, or whether charged surface grains are just ripped off and collisions of aggregates do not actually benefit from the charges on single dust grains at all.
There are also a number of experimental challenges, as dust aggregates are neither perfect spheres nor as monodisperse as commercial spherical submillimeter (submm) beads can often be. This can induce a charge bias to an ensemble of grains; for instance, all studied aggregates having the same polarity will inhibit clustering; not to mention that, in reality, dust comes with slightly different compositions, ranging from silicates to metallic iron, while the large internal surface is susceptible to water and conduction. 

In any case, after some years of experiments with monolithic particles, a further confirmation of the potential of tribocharging during planet formation could be obtained via the observation of charged dust aggregates and their clustering. This is the focus of this work. For the first time, we have observed tribocharging as well as clustering of dust aggregates in microgravity experiments at the drop tower in Bremen.

\section{Drop tower experiments}

\begin{figure}
   \centering
      \includegraphics[height=6.5cm]{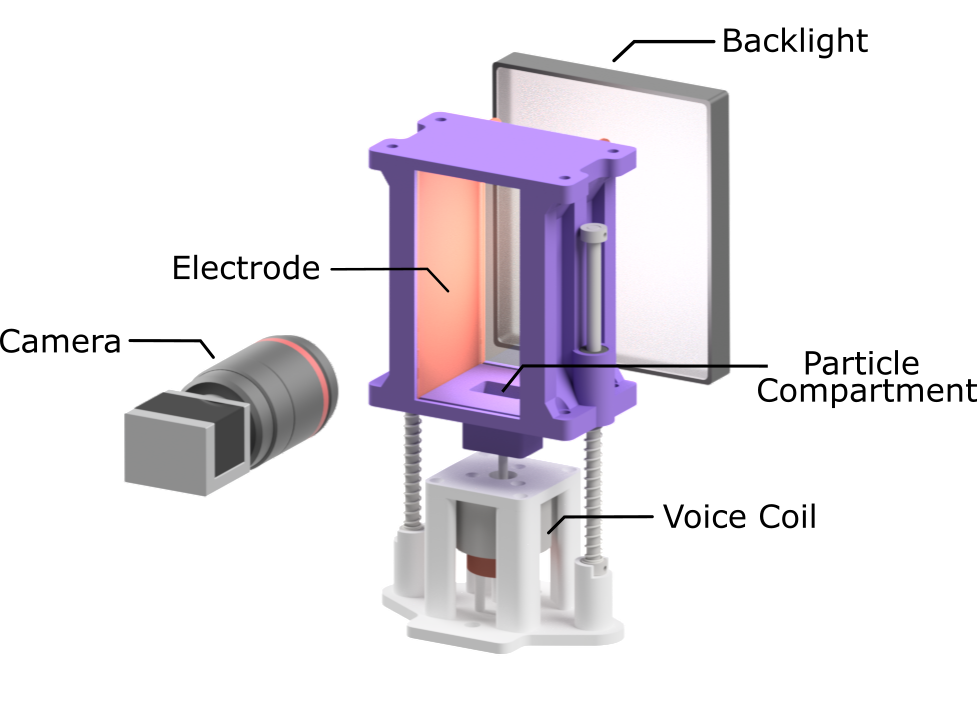}
      \caption{Basic setup of the drop tower experiments (taken from \citealt{Onyeagusi2024}).}
      \label{fig:aufbau}
\end{figure}

The basic setup of the experiment is shown in Fig. \ref{fig:aufbau}. The image is a copy taken from \cite{Onyeagusi2024} since it is essentially the same setup.

An ensemble of dust aggregates is placed in the particle compartment at the bottom of the 10\,cm $\times$ 5\,cm $\times$ 5\,cm experiment chamber. The chamber is vibrated on ground for 15\,min to (supposedly) electrically charge the sample. As the experiment is launched within the drop tower in Bremen, the aggregates are released into the chamber. Here, they can move freely in microgravity for about 9\,s. That means they can collide with each other or be attracted by the electrodes of a plate capacitor as an electric field is applied. This way, the motion, collisional outcomes, morphology, and net charges of the aggregates and clusters of aggregates can be determined. 
In contrast to the study of \cite{Onyeagusi2024}, the whole setup is placed within a vacuum chamber, and pressures of $\sim 10^{-2}$\,mbar are set for the experiments at hand. 

\subsection{Dust and aggregate sample}
The aggregates are prepared from a dust material that is also used as Martian simulant \citep[MGS;][]{Cannon2019}. The exact composition is of minor importance at this point. But since it consists in major part of silicates, we consider it a suitable first simulant for preplanetary dust.
The sample was milled to a grain size of a few µm. A size distribution of the grains is given in 
Fig. \ref{fig:sizes} (yellow). This size distribution was measured with a Mastersizer 3000.

The dust was vibrated on a commercial shaker, which led to the formation of aggregates.
The size of these aggregates is limited by bouncing quite similar to the bouncing barrier in protoplanetary disks. Thus, the particles simulate the bouncing barrier closely. The MGS sample has a bulk density of 1.29\,g/cm³ \citep{Cannon2019}. For porous aggregates, we determined a reduced density of 0.43\,g/cm³.
The aggregates were then sieved to yield an aggregate sample that is closer to being monodisperse within the regime of 150 -- 500\,µm. A more detailed size distribution of the aggregates is also shown in Fig. \ref{fig:sizes} in blue. This size distribution was taken from individual frames of the experiment's video material, as seen, for example, in Figs. \ref{fig:initial}, \ref{fig:speed}~(a),~and  \ref{fig:speed}(b). Particles that can be identified as single aggregates can be automatically traced to measure their sizes using the imaging software imageJ \citep{Schneider2012}.
The step of shaping the size distribution is not needed for the clustering but makes the experiments easier to carry out and analyze. {In the following, we will refer to the porous submm particles made from micron grains as ``aggregates,'' whereas we call stable conglomerates of these aggregates ``clusters'' or ``agglomerates.''

\begin{figure}
   \centering
      \includegraphics[width = \linewidth]{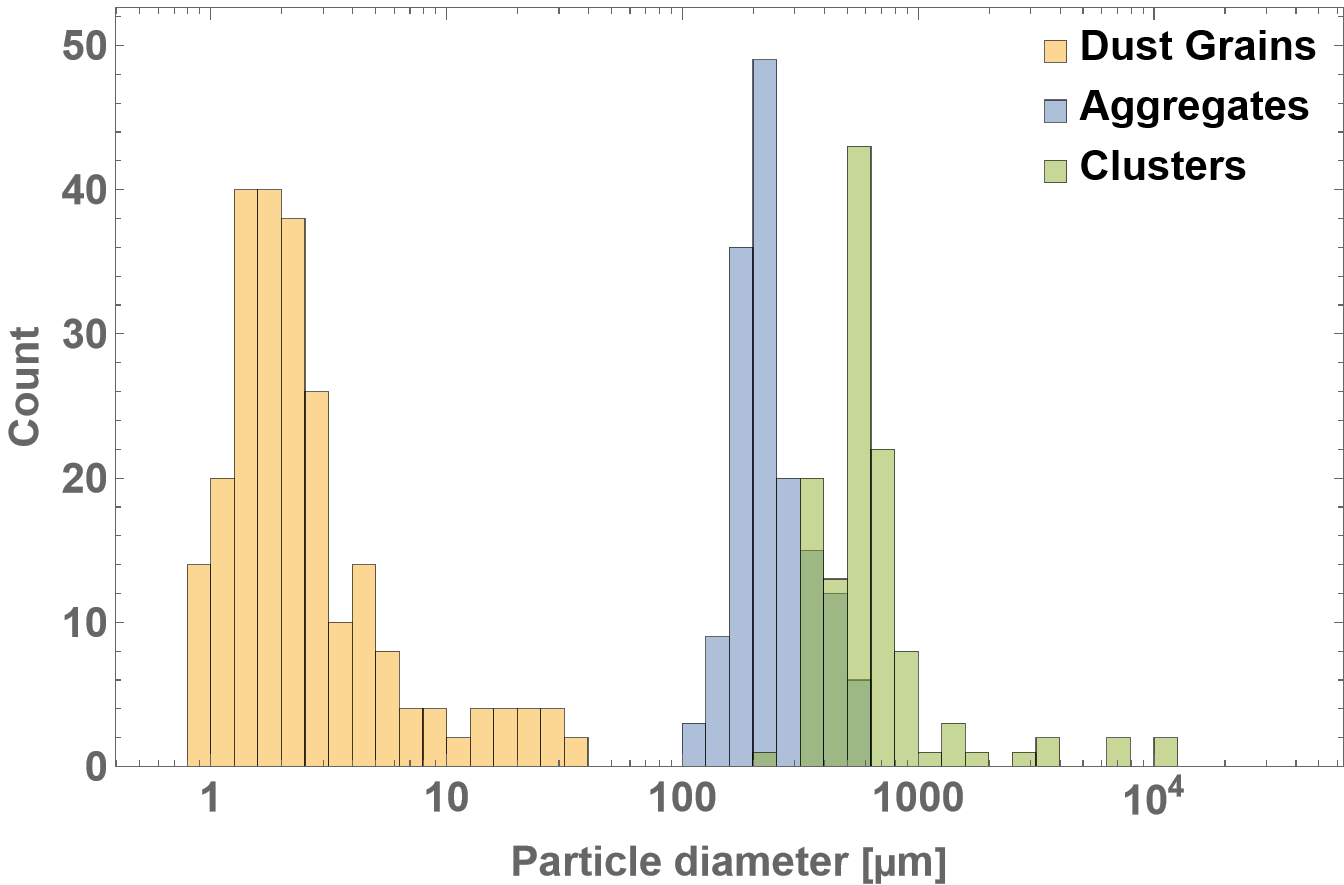}
      \caption{Size distributions of dust grains (yellow: small grains), dust aggregates (blue: medium size), and clusters of aggregates at a certain time in the experiment run (green). The cluster size distribution depends on the time of the experiment run and is not an equilibrium distribution. It shifts to larger values over time.}
      \label{fig:sizes}
   \end{figure}

Also, in contrast to the approach in \cite{Onyeagusi2024}, the aggregate sample was dried at 120\,$^{\circ}$C for 48\,h to remove most of the water within the pore space. Together with the fact that the experiments are carried out under vacuum conditions, this leaves only about a monolayer of water on the surfaces \citep{Steinpilz2019, Pillich2021, Kimura2015}.
Therefore, charging is still possible \citep{Becker2022}, but the conductivity of the surfaces does not have to be considered \citep{Becker2024}.

The particle motion was recorded with a spatial resolution of 32\,µm/px and a time resolution of 20\,ms. Initially, the particles are distributed rather homogeneously within the experiment chamber as shown in Fig. \ref{fig:initial}. We also included a microscope image of a dust aggregate as an inset in Fig. \ref{fig:initial} to highlight the fact that these particles seen in the black-and-white video frame are actual aggregates, and not individual monolithic grains. This is a crucial difference from earlier experiments.

\begin{figure}
   \centering
      \includegraphics[width = \linewidth]{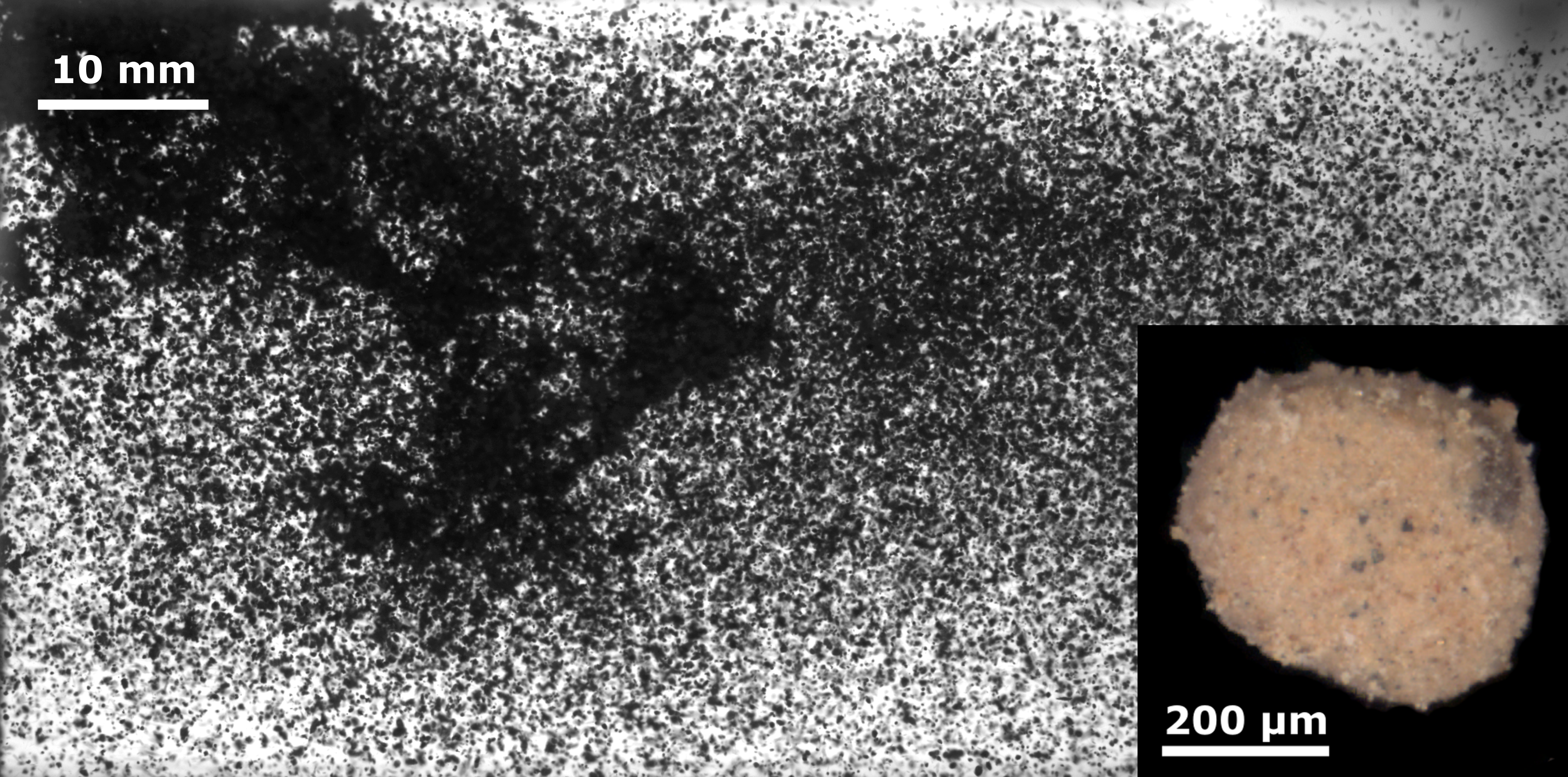}
      
      \caption{Dust aggregates after agitation of the sample in microgravity. Some denser regions are visible, but particles are generally spread out in a diffuse cloud from which clusters grow. We note that each individual particle is a dust aggregate. To emphasize this, a microscope image of the particles is shown.}
      \label{fig:initial}
   \end{figure}

\section{Results}

\subsection{Impact Velocities}

At the beginning of the experiment run, the chamber is vertically agitated to distribute the sample, as stated before. The chamber motion, however, has lateral components as well, which guide the particles to the center where they collide and build clusters. In Figs. \ref{fig:speed}(a) and \ref{fig:speed}(b), we can see that elongated structures form in the middle of the chamber, while there is a dearth of aggregates close to the top and lower wall. This accumulation increases over time. We use the distinction between areas of high particle density and low particle density to determine an average velocity with which aggregates move toward the center and collide. 

\begin{figure}
   \centering
   
            \includegraphics[width = \linewidth]{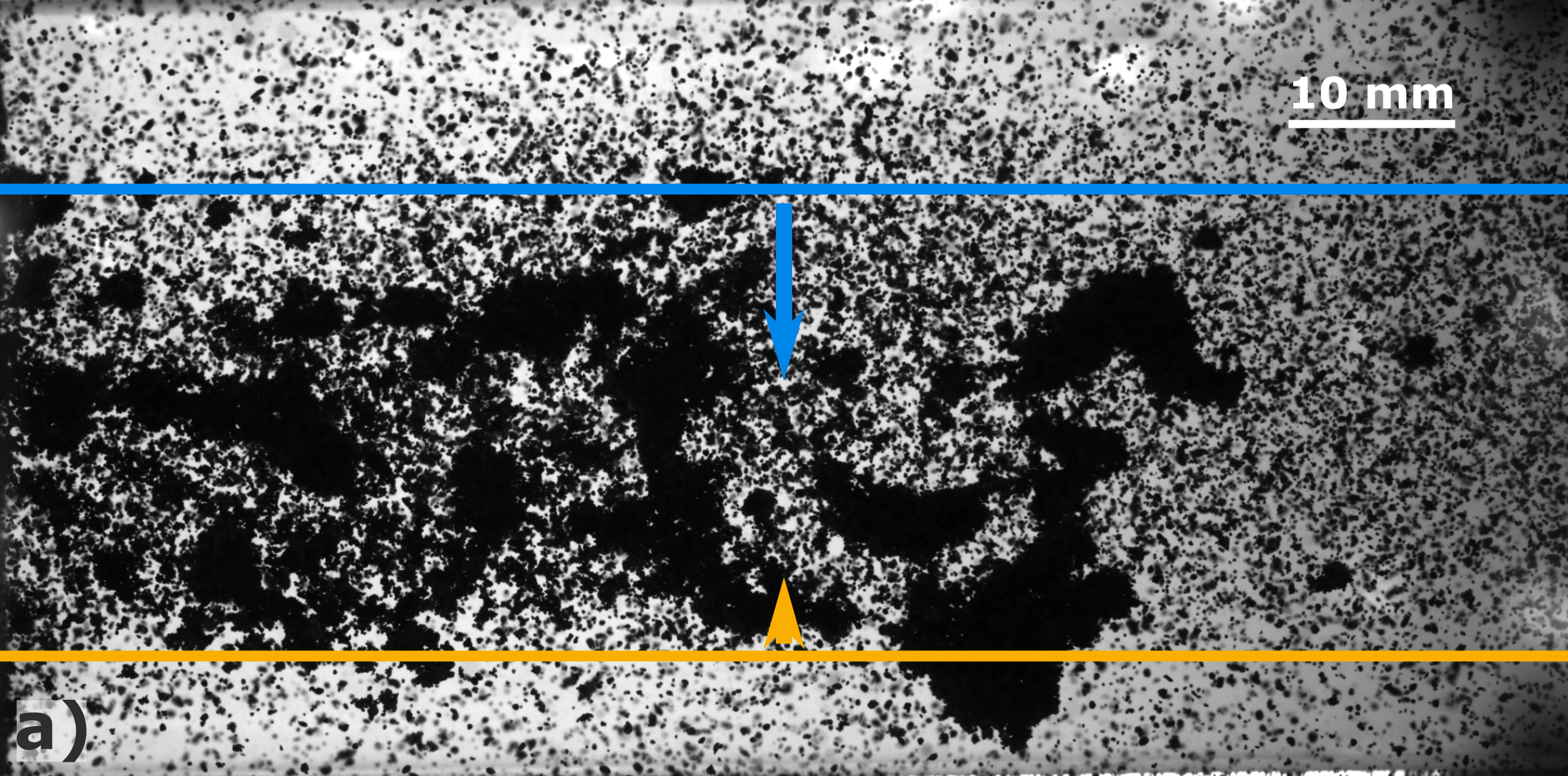}
        
            \includegraphics[width = \linewidth]{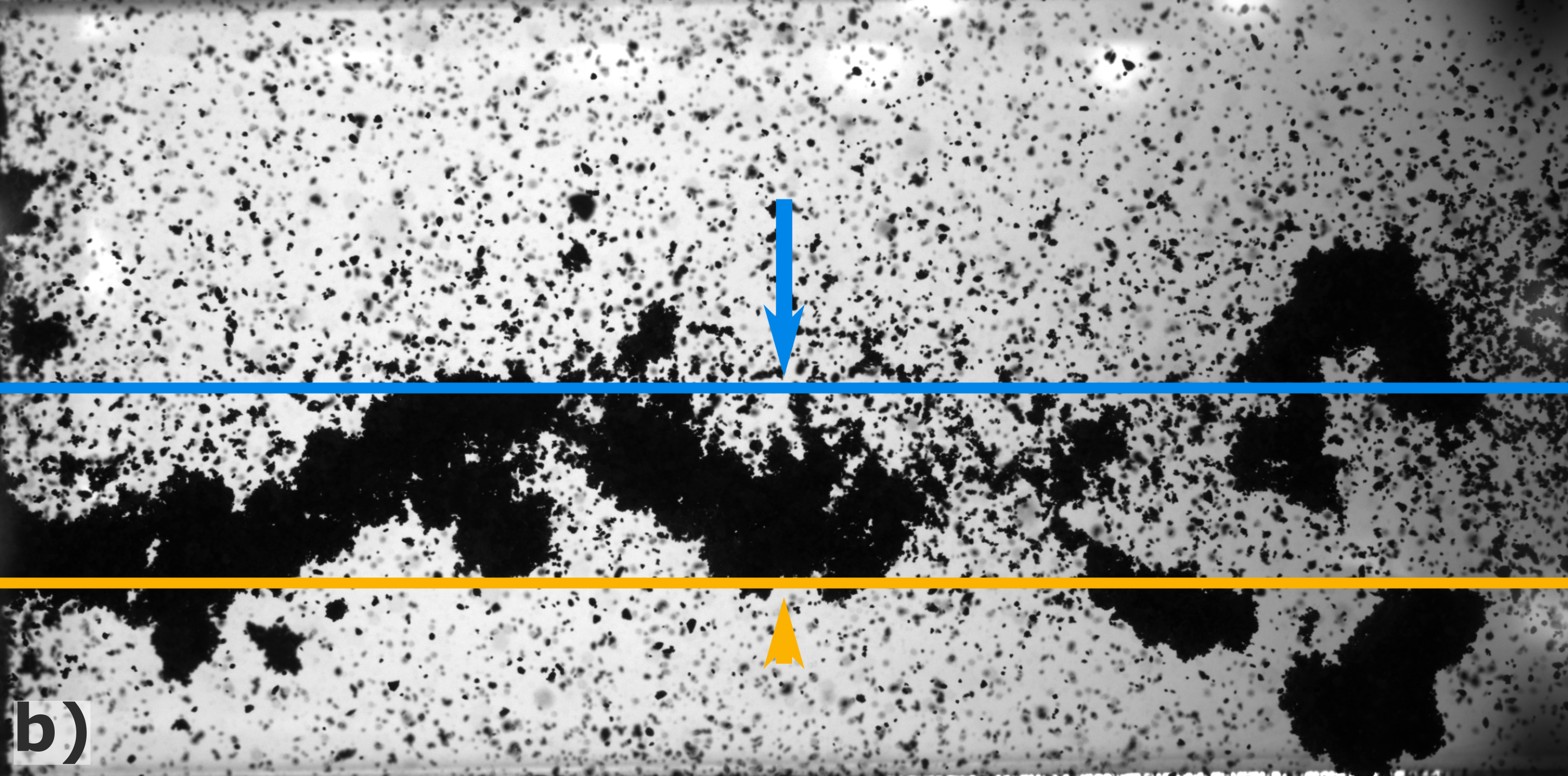}
        
            \includegraphics[width = \linewidth]{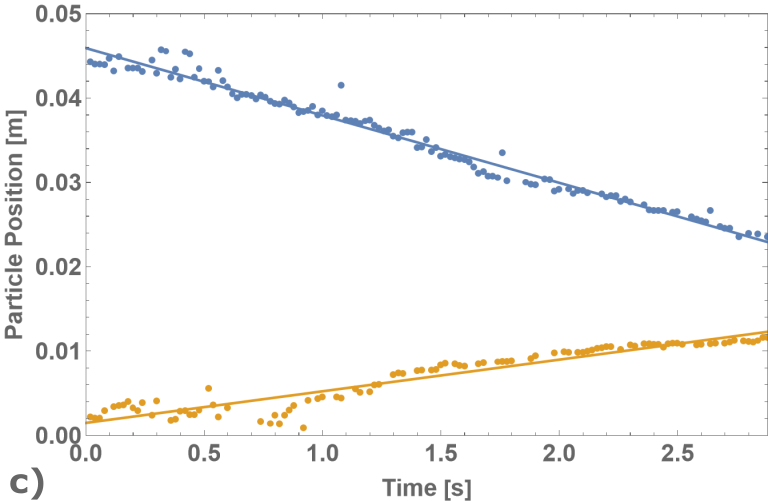}
      
      \caption{Particle movement in microgravity. Panels (a) and (b): Sequence of two images at different points in time with horizontal lines visualizing the systematic motion of the particles. Panel (c): A horizontal, average brightness threshold can be used to track the particle motion. The slopes of the particle positions give the typical collision velocities.}
      \label{fig:speed}
   \end{figure}
   
The particle motion can be approximated as a vertical movement from both sides to the center. In detail, the average horizontal brightness can be used to track the motion along a certain brightness threshold. This is visualized in Fig. \ref{fig:speed}(c), where the linear slopes are the velocities of the two approaching particle fronts. We find $11.5\,\times\,10^{-3}$\,m/s as an average relative velocity of the colliding particles. We use this approach due to the crowded images where individual trajectories of colliding aggregates are not easily tracked otherwise. It is only for some larger clusters that movements and collision velocities can be tracked directly. We did so for a small sample of clusters to check for consistency, but also found collision velocities on the order of 1\,cm/s. The current setup is not suitable for in-depth studies of collisions between clusters. Systematic studies of this require a dedicated setup, which is currently in development.
In any case, at the end of the microgravity time, the largest clusters are centimeter-sized as seen in Fig. \ref{fig:big}.

\begin{figure*}
   \centering
      \includegraphics[width = \textwidth]{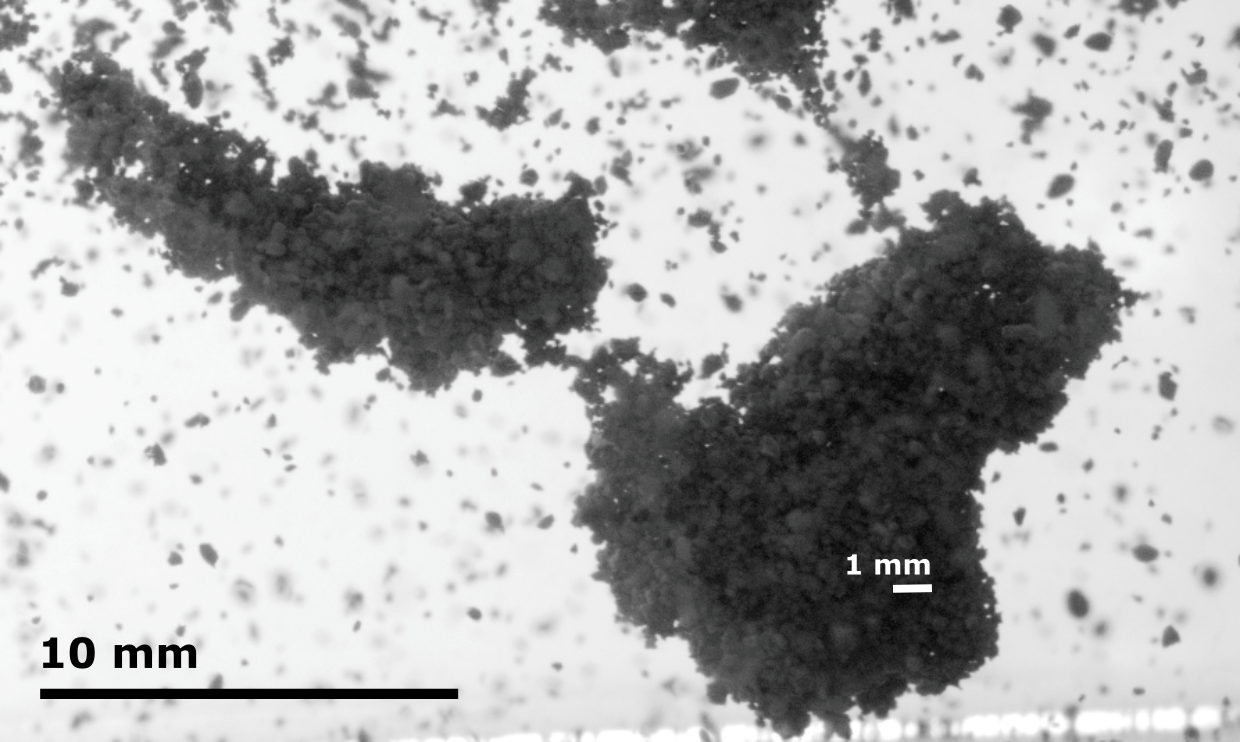}
      \caption{Contrast enhanced image of one of the largest clusters. The constituent aggregates are visible on a submm scale.}
      \label{fig:big}
\end{figure*}

We would not expect such large clusters to form at the given speed without electric charges. \citet{Teiser2021} carried out similar experiments with monolithic basalt beads at the lower end of the size range of our MGS aggregates and found that a charged sample formed compact clusters up to 5\,cm, while the uncharged sample only formed fractal mm-sized clusters. \citet{Onyeagusi2024} also carried out microgravity experiments with slightly bigger MGS aggregates (0.5 -- 2.0\,mm), but in that case, the sample was not heated beforehand, allowing for more water to be adsorbed to the surface and the porous interior. This made the sample slightly conductive and prevented charging. The authors observed hardly any clustering at all. Thus, we can conclude that electric charges on dusty aggregates amplify the stability of such clusters, analogously to the earlier findings for monolithic grains. To support this, the charge of the aggregates has to be determined. 

\subsection{Charges}
We measured the charge of the aggregates in another experiment run by applying a voltage of 400\,V to the capacitor plates.
Within the electric field, aggregates of different polarity are drawn to the two electrodes. In this way, we find charges of up to $10^5$\,e as seen in Fig. \ref{fig:charge}.

\begin{figure}
   \centering
      \includegraphics[width = \linewidth]{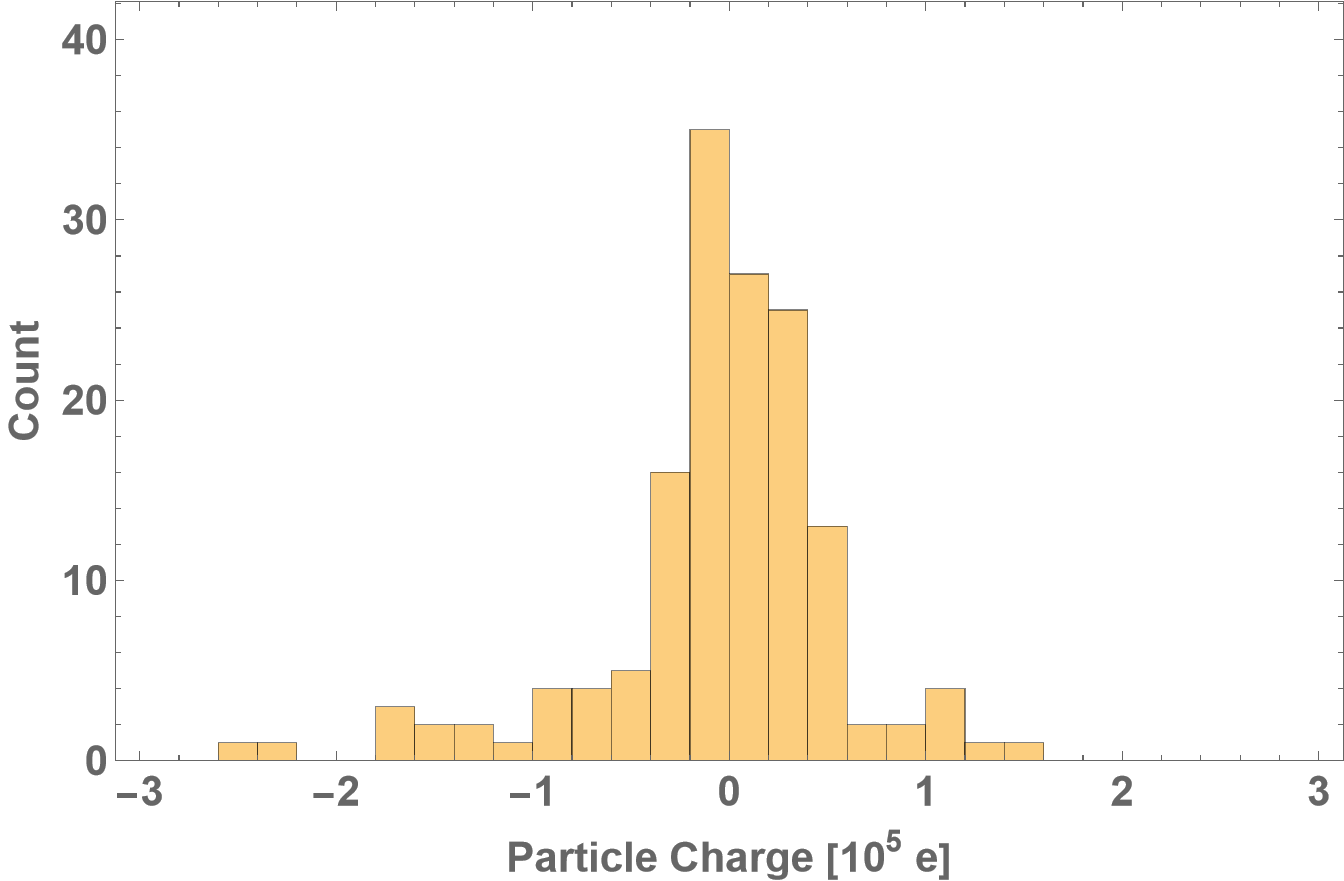}
      \caption{Charge distribution of dust aggregates.}
      \label{fig:charge}
   \end{figure}

The aggregate mass was estimated by taking an ellipsoid with an equivalent cross-section of a tracked particle and assuming a density of 0.43\,g/cm³. The charge can be estimated through the given acceleration of the mass within the electric field of 8\,kV/m. The charge distribution in Fig. \ref{fig:charge} includes only data from individual aggregates, not from the clusters. We observe charges of both polarities with no clear bias toward positive or negative. Considering the particle size in the order of 100\,µm, we get surface charges of up to $ 10^{-7}$\,C/m². This charge density is similar to earlier findings for monolithic glass or basalt spheres \citep{Jungmann2018, Steinpilz2020a, Jungmann2021, Onyeagusi2023b}. 

\section{Discussion}

We only see the growth of aggregates here with no signs of bouncing. Therefore, the particle velocity of 1\,cm/s is a lower limit for the maximum sticking velocity of aggregates and clusters thereof. The cm-clusters that form throughout the experiment have a mass of about $10^{-2}$\,g. This, for example, would be an order of magnitude larger than studied by \cite{Kothe2013} for presumably uncharged and fractal clusters. \cite{Brisset2016,Brisset2017} also studied dust aggregates consisting of mono- and polydisperse silicate grains. While the constituent µm-grains exhibit a small amount of charge, their aggregates were not sufficiently charged to show any effect in the experiment results. Their aggregates only formed fractal-like mm-sized clusters as well.

In comparison to these other works with a very similar setup, but mostly uncharged dust aggregates, we observed rather compact clusters right from the beginning. It seems to be a feature of charge stabilized clusters that they (re)-arrange in a compact way \citep{Steinpilz2020a, Teiser2021}. This most likely minimizes the electrostatic energy.

In previous works, we studied aggregates in a size range of 0.5 -- 2.0\,mm of the same material (MGS), but with different electrical properties. We conducted the same experiment within ambient air and without heating the dust sample to remove water adhered to the particles \citep{Onyeagusi2024}. This had a significant effect on the agglomerate stability in the face of an electric field, as the few layers of water that coated the surface (and presumably the inside of the porous aggregates) made the particles slightly conductive. The conductivity enabled charge separation within the existing small clusters as soon as an electric field was applied. Both opposing sides facing the electrodes would charge up, until the loosely bound aggregates disengaged from the agglomerate. In this way, the cluster was dissolved layer by layer. Additionally, this sample barely showed any clustering in the first place. We know that enhanced conductivity promotes internal discharging or charge equilibration. This is why we take the results from \citet{Onyeagusi2024} as a good means of comparison to a less-charged sample.

In summary, the electrical charge on single particles as well as dusty aggregates generated through tribocharging affects the process of particle growth. The borders between sticking, bouncing, and fragmentational collisions generally result from the velocity of the impacting particles and their mass, but if charge is added, these borders can be shifted, allowing for the formation of larger clusters. Electric charge not only facilitates sticking, it also creates clusters that are more compact, with strengthened inter-particle bonds opposing fragmentation; this is supported by the fact that we see more fractal-like particle growth for an uncharged sample \citep{Teiser2021}. The fact that this can not only be seen for idealized smooth, monolithic, and monodisperse samples -- but for dust aggregates as well -- strongly supports the claim that tribocharging plays a non-negligible role in planet formation.

\section{Conclusion}

Through microgravity experiments, we have observed that dust aggregates tribocharge and grow to larger clusters. At first sight, this is not different from what has been reported in earlier works. There is a subtle but important difference, however; in particular, the particles adopted in such prior studies were just monolithic beads or large single grains that clustered. In the present work, the particles are dust aggregates, adopted for the first time to simulate particles at the bouncing barrier. Although there might be quite a number of differences concerning tribocharging in detail, these aggregates form larger clusters easily, whereas neutral aggregates do not. This can be seen in comparisons to previous experiments \citep{Teiser2021, Onyeagusi2024}. This finding brings us one step closer to the reality of protoplanetary disk formation. It strongly suggests that tribocharging might be an important mechanism to glue dust aggregates together and shift the bouncing barrier to forge a sturdy bridge into a size range where hydrodynamic instabilities could potentially take over.

\begin{acknowledgements}
This project is supported by DLR Space Administration with funds provided by the Federal Ministry for Economic Affairs and Climate Action (BMWK) under grant numbers 50 WM 2142 and 50 WM 2442. 
\end{acknowledgements}

\bibliographystyle{aa} 
\bibliography{bib} 

\end{document}